\documentclass[11pt]{amsart}
\title {Tree Automata and Essential Input Variables}
\author{ Slavcho Shtrakov}
\address{Dept. of Computer  Sciences, South-West University,
 Blagoevgrad }
\email{shtrakov@aix.swu.bg}
\date{}
\usepackage{graphicx}
\usepackage{graphics}
\usepackage{latexsym}
\begin{document}
\newtheorem{l00}{\bf Lemma}
\newtheorem{p00}{\bf Procedure}
\newtheorem{t00}{\bf Theorem}
\newtheorem{e00}{\bf Example}

\newtheorem{d00}{\bf Definition}
\newtheorem{c00}{\bf Corollary}
\def\Pr{\bf Proof.\ }
\def\fbx{\hfill${}^{\rule{2mm}{2mm}}$}
\def\la{\leftarrow}
\def\ra{\rightarrow}
\def\Ra{\Rightarrow}
\def\F{{ F}}
\def\A{{ A}}
\def\La{\Leftarrow}
\relax
\begin{abstract}
We introduce the essential inputs (variables) for terms (trees) and
tree automata. It is proved that if an input $x_i$ is essential for
a tree $t$ and an automaton $A$ then there is a chain of subtrees
connecting  $x_i$ with the root of $t$ such that $x_i$ is essential
for each subtree belonging to this chain. There are investigations
which treat some rules for removing and adding of fictive
(non-essential) inputs of a term. We consider a new point of view of
minimization of tree-automata and tree-languages. Such minimization
is realized by a procedure (algorithm).
\end{abstract}
\maketitle
 \noindent {\it AMS, subject classification}: 03D05,
68Q70, 03D15, 06B25
\\
{\it Key words and phrases}: Tree, Tree Automata, Essential Input.
\section{Introduction}
The consideration that finite automata may be viewed as unary
algebras is attributed to J.B\"uchi and J.Wright (1960). In  many
papers trees were defined as terms. Investigations on regular and
context-free tree grammars dated back to the  60's.\\
Tree automata are designed in context of circuit verification and
logic programming.In the 70's some new results were obtained
concerning tree automata, as an important part of the theoretical
basis of computing and programming. So, since the end of the 70's
tree automata have been used as powerful tools in program
verification. There are many results connecting properties of
programs or type systems or rewrite systems with automata  (see e.g.
\cite{Com}).
\\
 The algebraic theory of terms was created and developed to the
equational theory in the work of A.Malc'ev and  G.Gr\"atzer (see
\cite{Mal,Gr}). There are many new results concerning
hypersubstitutions, hyperidentities, solid varieties, term (tree)
algebra (\cite {Den3,Ros}).
\\ The theory of essential variables for discrete functions was
developed by  S.Jablonsky, A.Salomaa, K.Chimev and others
(\cite{Ch,Jab,Sa}). Discrete functions on a finite domain can be
viewed as elements of a term algebra. The results obtained here are
very useful for analysis and synthesis of functional schemes and
circuits.
\\
The present paper is an attempt to connect these three fields of
theoretical computer science.
\section{Basic Definitions and Notations}
Let ${F}$ be any finite set, the elements of which are called $ operation\
symbols. $ Let $\tau:{{F}}\to N$ be a mapping into the non negative integers;
for $f\in{ F},$ the number $\tau(f)$ will denote the {\it arity } of the
operation symbol $f.$ The pair $(F,\tau)$ is called {\it type} or
{\it signature}.
 Often if it is obvious what the set ${F}$ is,
we will write "$ type\ \tau$". The set of symbols of arity
$p$ is denoted by ${ F}_p.$ Elements of arity $0,1,\ldots, p$
respectively are called {\it constants(nullary), unary,...,$p$-ary} symbols.
We assume that ${ F}_0\neq\emptyset.$
\begin{d00}
\rm\label{d2}
Let $X_n=\{x_1,\ldots,x_n\}, n\geq 1,$ be a set of variables with
$X_n\cap{{F}}=\emptyset.$
The set $W_{\tau}(X_n)$ of {\it  $n-$ary terms\ of\ type} $\tau$
with\ variables\ from\ $ X_n $ is defined as the smallest
set for which:
\\
$(i)$\ $F_0\subseteq W_{\tau}(X_n)$ and
\\
$(ii)$\ $X_n\subseteq W_{\tau}(X_n)$ and
\\
$(iii)$\ if $p\geq 1, f\in F_p $ and
$t_1,\ldots,t_p\in W_{\tau}(X_n)$ then $f(t_1,\ldots,t_p)\in W_{\tau}(X_n).$
\end{d00}
By $W_\tau(X)$ we denote the following set
$$W_\tau(X):=\cup_{n=1}^{\infty} W_\tau(X_n),$$
where $X=\{x_1,x_2,\ldots \}.$
\\
If $X=\emptyset$ then $W_\tau(X)$ is also written $W_\tau.$ Terms in
$W_\tau$ are called {\it ground terms.}
\\
Let $t$ be a term. By $Var(t)$  the set of all variables from
$X$ which occur in $t$ is denoted. The elements of $Var(t)$ are  called  {\it input variables} for $t$.
 \\
Let $t$ be a term and suppose we are given a term $s_x$ for every $x\in X.$
The term denoted by $t(x\la s_x),$ is obtained by substituting in $t,$
simultaneously for every $x\in X,\quad s_x$ for each occurrence of $x.$
The formal definition by term induction reads as follows:
\\
$(i)$\quad if $t=x\in X,$ then $t(x\la s_x)=s_x;$
\\
$(ii)$\quad if $t=f_0\in { F}_0,$ then $t(x\la s_x)=f_0;$
\\
$(iii)$\quad if $t=f(t_1,\ldots,t_n),$ then
$t(x\la s_x)=f(t_1(x\la s_x),\ldots,t_n(x\la s_x)).$
\\
If  $t,s_x\in W_{\tau}(X)$  then $t(x\la s_x)\in W_{\tau}(X).$
\\
If $t,s_x\in W_\tau(X_n)$, one may then write $t(x\la s_x)$ in the more
explicit form $t(x_1\la s_{x_1},\ldots,x_n\la s_{x_n}).$
\\
Any subset $L$ of $W_{\tau}(X)$ is called {\it term-language}
or {\it tree-language.}
\begin{d00} \rm\label{d3}
Let $t$ be a term of type $\tau.$ We define the $depth$ of $t$
in the  following inductive way:\\
$(i)$ if $t\in X\cup F_0$ then $Depth(t)=0;$\\
$(ii)$ if $t=f(t_1,\ldots,t_{n})$ then
$Depth(t)=max \{Depth(t_1),\ldots,Depth(t_{n})\} +1.$
\end{d00}
 The {\it tree} of a term  $t$ is defined as follows:
\\
$(i)$\quad if $t=x_k$ (or $t=f, f\in{ F}_0$) then
the tree of the term $t$ consists of one node
labeled with $x_k$ (or $f$ respectively)
and this node is the root of the tree;\\
$(ii)$\quad if $t=f(t_1,\ldots,t_n)$ then the tree of $t$ has as root a node
labeled with $f$ and its successors are the roots of the terms
$t_1,\ldots,t_n.$
\\
Often, when we write "term $t$" we will
mean the corresponding
tree and conversely.
\par
Let $N$ be the set of natural numbers and $N^*$ be the set of finite strings over $N.$  The set $N^*$ is naturally ordered by
$n\preceq m \iff n$\ is a prefix of $m.$
\par
Now, a finite ordered tree (term) $t$ over a set of operation symbols (labels) $ F$ is
a mapping from a prefix-closed set $Pos(t)\subseteq N^*$ into $ F.$
Thus a term $t\in W_{\tau}(X)$ may be viewed as a finite ordered tree, the leaves of
which are labeled with variables or constant symbols
and the internal nodes are labeled with operation symbols of
positive arity, with out-degree equal to the arity of the label,
i.e. a term $t\in W_{\tau}(X)$ can also be defined as a partial function
$t:N^*\to {{F}}\cup X$ with domain $Pos(t)$ satisfying the following properties:
\\
$(i)$\quad $Pos(t)$ is nonempty and prefix-closed;
\\
$(ii)$\quad For each $p\in Pos(t)$, if $t(p)\in { F}_n,$
$n\geq 1$
then $\{i|pi\in Pos(t)\}=
\{1,\ldots,n\};$
\\
$(iii)$\quad For each $p\in Pos(t)$, if $t(p)\in X\cup F_0$
then $\{i|pi\in Pos(t)\}=\emptyset.$
\\
\par
The elements of $Pos(t)$ are called  {\it positions.} A {\it frontier position} is a position
$p$ such that
$\forall \alpha\in N,\quad p\alpha\notin Pos(t).$ The set of frontier positions is denoted by
$FPos(t).$ Each position $p$ in $t$ with $t(p)\in X$ is called {\it variable position}.
The set of variable positions of $t$ is denoted by $VPos(t).$ Clearly
$VPos(t)\subseteq FPos(t).$ The elements of the set $CPos(t)=FPos(t)\setminus VPos(t)$
are caled {\it constant positions.}
\par
A {\it subterm} $t|_p$ of a term $t\in W_{\tau}(X)$ at position $p$ is
defined as follows:
\\
$(i)$\quad $Pos(t|_p)=\{i|pi\in Pos(t)\};$
\\
$(ii)$\quad $\forall j\in Pos(t|_p),\quad t|_p(j)=t(pj).$
\\
The subtrees at the frontier positions for $t$ are called {\it
inputs} of $t.$
\par
 By $t[u]_p$ we  denote the term obtained by replacing the subterm $t|_p$ in $t$
 by $u.$
\\
 We write $Head(t)=f$ if and only if $t(\varepsilon)=f$, where
$\varepsilon$ is the empty string in $N^*,$
i.e. $f$ is the {\it
root symbol} of $t.$
\\
Thus we define a partial order relation in the set of all terms $W_{\tau}(X).$
We denote by $\unlhd$ the subterm ordering, i.e. we write $t\unlhd t'$ if
there is a position $p$ for $t'$ such that $t=t'|_p$ and one says that
$t$ is a subterm of $t'.$ We write $t\lhd t'$ if $t\unlhd t'$ and
$t\neq t'.$
\\
A chain of subterms $Ch:=t_{p_1}\lhd t_{p_2}\lhd \ldots \lhd t_{p_k}$ is called {\it strong}
if for all $j\in \{1,\ldots,k-1\}$
there does not exist a term $s$ such that
$t_{p_j}\lhd s\lhd  t_{p_{j+1}}.$
\section{Finite Tree Automata  and
Essential Variables}
\begin{d00}\rm\label{d5}
A  FTA  is a tuple
$\A=\langle Q,\F, Q_f,\Delta\rangle$ where:
\\
- $Q$ is a finite set of states;
\\
- $Q_f\subseteq Q$ is a set of final states;
\\
- $\Delta$ is a set of transition rules i.e. if
$$\F= \F_0\cup\F_1\cup\ldots\cup\F_n\quad
{\mbox{\rm then}}\quad \Delta=\{\Delta_0,\Delta_1,\ldots,\Delta_n\},$$
where  $\Delta_i$
are  mappings
$\Delta_0:F_0\ra Q,$
and
$\Delta_i:\F_i\times Q^i\ra Q,$ for $ i=1,\ldots,n.$
\end{d00}
\noindent
We will suppose that $\A$ is complete i.e. the $\Delta$'s are total mappings on their
domains.
\\
Let $Y\subseteq X$ be a set of variables and
$\gamma:Y\ra \F_0$
be a function which assigns  nullary operation symbols (constants) to each input
variable from $Y.$
 The function $\gamma$ is called
{\it assignment on the set of inputs $Y$} and the set of such assignments
 will be denoted by
$Ass(Y,\F_0).$
\\
Let $t\in W_{\tau}(X),$ $\gamma\in Ass(Y,\F_0)$ and
$Y=\{x_1,\ldots,x_m\}.$ By $\gamma(t)$ the term
$\gamma(t)=t(x_1\la \gamma(x_1),\ldots,x_m\la \gamma(x_m))$
will be denoted.
 \\
So, each assignment $\gamma\in Ass(Y,\F_0)$ can be extended to
a mapping defined on the set
$W_{\tau}(X)$ of all terms.
\\
Let $t\in W_{\tau}(X),$ and $ \gamma\in Ass(X,\F_0).$
The automaton
$\A=\langle Q,\F, Q_f,\Delta\rangle$
 runs over $t$ and $\gamma.$ It starts at
leaves of $t$ and moves downwards, associating along a run a
resulting state with each subterm inductively:
\\
$(i)$\quad If $Depth(t)=0$ then the automaton $\A$ associates with $t$ the
state $q\in Q,$ where
$$q=\left\{\begin{array}{ll}
       \Delta_0(\gamma(x_i))  & \quad \mbox{\rm if}\quad t=x_i\in X;\\
       \Delta_0(f_0)  & \quad \mbox{\rm if}\quad t=f_0\in\F_0.
        \end{array}
        \right.$$
$(ii)$\quad Let $Depth(t)\geq 1.$
If $t=f(t_1,\ldots,t_n)$ and the states  $q_1,\ldots,q_n$ are
associated with the subterms(subtrees) $t_1,\ldots,t_n$ then with $t$
the automaton $\A$ associates the state $q,$ where
$q=\Delta_n(f,q_1,\ldots,q_n).$
\\
Note that the automaton runs only over ground terms and each
assignment from $Ass(X,F_0)$ transforms any tree as a ground term.
\\
The initial states are the states associated with the leaves of the tree
as for terms with depth equals to 0 i.e. as in the case $(i).$
\\
A term $t\ ,t\in W_{\tau}(X)$ is accepted by a tree automaton
$\A=\langle Q,\F, Q_f,\Delta\rangle$ if there exists an assignment
$\gamma$ such that when running over $t$ and $\gamma$ the automaton
$\A$ associates with $t$ a final state  $q\in Q_f.$
\\
When $\A$ associates the state $q$ with a subterm $s,$ we will write
$ \A(\gamma,s)=q.$
\\
Let $t\in W_{\tau}(X)$ be a term and $\A$ be a tree automaton which
accepts $t.$ In this case one says that $\A$ {\it recognizes} $t$ or
$t$ is {\it recognizable} by $\A.$ The set of all by $\A$
 recognizable terms is called {\it tree-language} recognized by $\A$ and will be denoted
by $L(\A).$

\begin{d00}\rm\label{d6}
Let $t\in W_{\tau}(X)$ and let $A$ be a tree automaton. An input
variable $x_i\in Var(t)$ is called {\it essential} for the pair
$(t,{ A})$ if there exist  two assignments $\gamma_1, \gamma_2\in
Ass(X,{ F}_0)$ such that
$$\gamma_1(x_i)\neq \gamma_2(x_i),\quad
\forall x_j\in X,\ j\neq i \quad \gamma_1(x_j)=\gamma_2(x_j)$$
 with
${ A}(\gamma_1,t)\neq { A}(\gamma_2,t)$
i.e. ${ A}$ stops in
different states when running over $t$ with  $\gamma_1$ and  with
$\gamma_2.$
\end{d00}
The set of all essential inputs for $(t,{ A})$ is denoted by
$Ess(t,{ A}).$ The input variables from
$Var(t)\setminus Ess(t,{ A})$
are called {\it fictive } for  $(t,{ A}).$
\begin{t00}\label{t1}\rm
If $x_i\in Ess(t,{ A})$ then there exists a strong
chain
$x_i=t_1\lhd t_2\lhd \ldots \lhd t_k\unlhd t$
such that $x_i\in Ess(t_j,{ A})$ for $j=1,\ldots,k.$
\end{t00}
\Pr\rm
Let $t\in W_{\tau}(X)$ and $\gamma_1,\gamma_2\in Ass(X,{ F}_0)$ be
a term and two assignments, such that
$\gamma_1(x_i)\neq\gamma_2(x_i)$ and $\gamma_1(x_j)=\gamma_2(x_j),\quad
\mbox{\rm for}\quad j\neq i $ with
$${ A}(\gamma_1,t)\neq { A}(\gamma_2,t).$$
At first, if $Depth(t)=1$ then the chain $x_i\unlhd t$ is strong and the
theorem is proved in this case.
\\
Secondly, let us assume $Depth(t)\geq 2$ and
$t=f(t_1,\ldots,t_n).$ Suppose that the theorem is true
for the subterms $t_1,\ldots,t_n$ i.e. if
$x_i\in Ess(t_j,A)$ then there exists at least one
strong chain
$x_i=t_{p_1}\lhd t_{p_2}\lhd \ldots \lhd t_{p_k}\unlhd t_j$ with
$x_i\in Ess(t_{p_l},A),$ $l\in \{1,\ldots,k\},$ and
$j\in \{1,\ldots,n\}.$
\\
It is sufficient to prove that $x_i\in Ess(t_j,A)$ for at least one
$j,$ $j\in\{1,\ldots,n\}.$
\\
Suppose that $x_i\notin Ess(t_j,{ A})$
for all $j,\quad j\in\{1,\ldots,n\}.$
This implies that ${ A}(\gamma_1,t_j)={ A}(\gamma_2,t_j)$
for all $j,\quad j\in\{1,\ldots,n\}.$
Let us calculate  ${ A}(\gamma_1,t)$   and ${ A}(\gamma_2,t).$
$${ A}(\gamma_1,t)=
\Delta_n(f,{ A}(\gamma_1,t_1),
\ldots,{ A}(\gamma_1,t_n))=$$
$$=
\Delta_n(f,{ A}(\gamma_2,t_1),
\ldots,{ A}(\gamma_2,t_n))={ A}(\gamma_2,t).$$
This contradicts  ${ A}(\gamma_1,t)\neq { A}(\gamma_2,t).$
 Hence there exists a
subterm $t_j,\quad j\in\{1,\ldots,n\}$ of $t$
 such that $x_i\in Ess(t_j,{ A}).$
\fbx
\par
It is easy to see that if \ $\forall \gamma\in Ass(X,{ F}_0)\quad
 { A}(\gamma,t')={ A}(\gamma,t) $
then
$$Ess(t,{ A})=Ess(t',{ A}).$$
\section{Removing and Adding of Fictive Inputs }
In this section we consider two types of changing the trees recognized by an
automaton. The first one leads to a simplification of the trees and the second one
increases the complexity of trees.
\subsection{Removing of Fictive Inputs (RFI) }\label{r}
We consider two types of removing rules over a tree and an automaton.
\\
$(i)$  \quad
Let $q_0\in Q$ with $\Delta_0(f_0)=q_0,\quad f_0\in\F_0.$
Let $p_1\in Pos(t)$ be
a variable position for $t,$ labeled by $x_i.$ There is a unique strong chain
$x_i=t|_{p_1}\lhd \ldots\lhd t|_{p_k}=t$
which connects the leaf $t|_{p_1}$ and the root of $t.$
If there is a subtree $t|_{p_j}$ of $t$ from this chain with
$x_i\notin Ess(t|_{p_j},A) $ then substitute in $t$ the term $t|_{p_j}(x_i\la f_0)$
instead of $t|_{p_j}.$
\\
 $(ii)$\quad If $t_1\lhd t_2\unlhd t$ and
 \begin{center}
$\forall \gamma\in Ass(X,\F_0)\quad \A(\gamma,t_1)=
\A(\gamma,t_2)$
\end{center}
then we remove the subtree $t_2$ and put the subtree $t_1$ instead of
$t_2.$
\\
Clearly the rules $(i)$ and $(ii)$ lead to simplify the trees.
\subsection{Adding of Fictive Inputs (AFI)} \label{a}
\par
There are two rules to add fictive inputs which correspond
to the two RFI-rules.
The first one treats the case
when we want to add a simple input variable and the second one is for addition
of a term at the place of a fictive input variable.
\\
$(i)$  \quad
Let $p_1\in Pos(t)$ be
a constant position for $t,$ labeled by $f_0.$ There is a unique strong chain
$f_0=t|_{p_1}\lhd \ldots\lhd t|_{p_k}=t$
which connects the leaf $t|_{p_1}$ and the root of $t.$ Let $x_i\in X.$
If there is a subtree $t|_{p_j}$ of $t$ from this chain with
$x_i\notin Ess(t|_{p_j}[x_i]_{p_1},A) $ then substitute in $t$ the term $t_{p_j}[x_i]_{p_1}$
instead of $t|_{p_j}.$
\\
 $(ii)$\quad If $t_1\lhd t_2\unlhd t$ and
 \begin{center}
$\forall \gamma\in Ass(X,\F_0)\quad \A(\gamma,t_1)=
\A(\gamma,t_2)$
\end{center}
then substitute in $t$  the term $t_2$ instead of
$t_1.$
\par
When a term $t'$ is obtained from $t$ by some RFI-rule we will denote this
by $t\vdash_R t'$ and if there are terms $t_1,\ldots,t_k$ with
$t\vdash_R t_1\vdash_R\ldots\vdash_R t_{k-1}\vdash_R t_k=t'$ then $t'$
 is called
{\it ${ A}$-reduction} of $t$ and we will use the  denotation
$t\models_R t'.$
\\
When $t'$ is a resulting term
under  some AFI-rule over $t$ it is denoted
by $t\vdash_A t'$ and if there are terms $t_1,\ldots,t_k$ with
$t\vdash_A t_1\vdash_A\ldots\vdash_A t_{k-1}\vdash_A t_k=t'$ then $t'$
is called
{\it ${ A}$-extension} of $t$ and we will use the denotation
$t\models_A t'.$
\\
It is no difficult to see that if $t$ and $s$ are two terms then $t\models_R s \iff s\models_A t.$
\begin{l00}\label{l3}\rm
Let $t=f(t_1,\ldots,t_n)$ and $s=g(s_1,\ldots,s_m)$ be two terms.
If $t\models_R s\ (t\models_A s)$ then
$\forall i\in\{1,\ldots,m\}, \exists j\in\{1,\ldots,n\} $ such that
\\
$t_j\models_R s_i\ (t_j\models_A s_i).$
\end{l00}
\begin{d00}\rm\label{d9}
Two terms $t$ and $s$ are called {\it ${ A}$-equivalent}
$(t\simeq_{ A} s)$
iff
$$\forall  \gamma\in Ass(X,{ F}_0)\quad
{ A}(\gamma,t)=
{ A}(\gamma,s).$$
\end{d00}
Thus  $\simeq_{\A}$ is an equivalence relation, i.e.
\\
$(i)$\quad $\forall t\in W_{\tau}(X)\quad t\simeq_{ A} t;$
\\
$(ii)$\quad $\forall t,s\in W_{\tau}(X)\quad t\simeq_{ A} s\Rightarrow
s\simeq_{ A} t;$
\\
$(iii)$\quad $\forall t,s,r\in W_{\tau}(X)\quad
(t\simeq_{ A} s\quad \& \quad s\simeq_{ A} r)\Rightarrow
t\simeq_{ A} r.$
\par
A term $t\in W_{\tau}(X)$ is called {\it $F_0-$covered} w.r.t. the
automaton $A$ if
$$\forall\ \gamma\in Ass(X,F_0)\quad \exists\ f_0\in F_0 \quad
A(\gamma,t)=\Delta_0(f_0).$$
\begin{t00}\label{tn1}\rm
Let $t,s\in W_{\tau}(X)$ and $s$ be a $F_0-$covered term w.r.t. $A.$
If ${p_1}$ is a variable position for $t, $ labeled by $x_i$ and there
is a prefix $p_j$ of $p_1$ with $x_i\in Var(t|_{p_j})\setminus Ess(t|_{p_j},A)$
then
$$\forall \ \gamma\in Ass(X,F_0)\quad
A(\gamma,t)=A(\gamma,t[s]_{p_1}).$$
\Pr\rm
At first let $Depth(t)=1$ (note that the case $Depth(t)=0$ is trivial).
Without loss of generality let us suppose $p_1=i$ and
\\
 $t=f(y_1,\ldots,y_{i-1},x_i,y_{i+1},\ldots,y_n),$ $y_j\in X\cup F_0$
for
\\
$j\in \{1,\ldots,i-1,i+1,\ldots,n\}$ and $x_i\in X.$
Clearly $x_i\in Var(t)\setminus Ess(t,A).$
Let $\gamma\in Ass(X,F_0).$ Consider the term
$v=t[s]_{p_1}=f(y_1,\ldots,y_{i-1},s,y_{i+1},\ldots,y_n).$
\\
Suppose $A(\gamma ,t)\not= A(\gamma ,v).$
Let us set $q_1=A(\gamma ,s),\quad q_2=\Delta_0(\gamma(x_i)).$
The supposition implies $q_1\not= q_2.$ On the other side $s$ is
$F_0-$covered and there is $f_0\in F_0$ such that $\Delta_0(f_0)=q_1.$
Let us consider the following assignment:
\[ \gamma_1(x)=\left\{\begin{array}{lll}
        \gamma(x) & \mbox{\rm if} & x\not= x_i;\\
        f_0 & \mbox{\rm if} & x=x_i.
\end{array}
\right.\]
It is easy to see that
$A(\gamma_1 ,t)=A(\gamma ,t[s]_{p_1})=A(\gamma ,v).$
Hence $A(\gamma_1 ,t)\not=A(\gamma ,t)$ and $x_i\in Ess(t,A),$
a contradiction.
\\
Secondly, let $Depth(t)\geq 2.$ Then $t=f(t_1,\ldots,t_n)$
where $t_j\in W_{\tau}(X)$ for $j\in\{1,\ldots,n\}.$
 Suppose the theorem is valid for
the terms $t_1,\ldots,t_n$ i.e. if
$x_i=t|_{p_1}\lhd\ldots\lhd t|_{p_k}\unlhd t_j$ is a strong chain with
$x_i\in Var(t|_{p_l})\setminus Ess(t|_{p_l},A)$
for some prefix $p_l$ of $p_1$ then
$$\forall \ \gamma\in Ass(X,F_0)\quad
A(\gamma,t_j)=A(\gamma,t_j[s]_{p_1}).$$
This equation
and $t[s]_{p_1}=f(t_1,\ldots,t_{j-1},t_j[s]_{p_1},t_{j+1},\ldots,t_n)$
imply
$$\forall \ \gamma\in Ass(X,F_0)\quad
A(\gamma,t)=A(\gamma,t[s]_{p_1})$$
since
$x_i=t|_{p_1}\lhd\ldots\unlhd t|_{p_k}\unlhd t_j \unlhd t$ is the unique strong
chain connecting $t|_{p_1}$ and the root of $t.$
\end{t00}
\fbx
\begin{e00}\label{e1}\rm
Let ${A}=\langle Q,{ F}, Q_f,\Delta\rangle$  with
\\
${ F}_0=\{0,1\}$, ${ F}_1=\{f_1\}$, ${ F}_2=\{g_1,g_2\},$
$Q=\{q_0,q_1\}$, $Q_f=\{q_1\}$,
\\
$\Delta_0(0)=q_0$, $\Delta_0(1)=q_1$,
$\Delta_1(f_1,q_0)=q_1$,
$\Delta_1(f_1,q_1)=q_0$,
\\
$\Delta_2(g_1,q_0,q_1)=
\Delta_2(g_1,q_1,q_0)=
\Delta_2(g_1,q_1,q_1)=q_1$,
$\Delta_2(g_1,q_0,q_0)=q_0$,
\\
$\Delta_2(g_2,q_0,q_0)=\Delta_2(g_2,q_0,q_1)=
\Delta_2(g_2,q_1,q_0)=q_0$,
$\Delta_2(g_2,q_1,q_1)=q_1.$
\\
Let us consider the term
$t=g_1(g_2(g_1(f_1(x_3),x_2),x_2),g_1(x_1,g_2(x_1,f_1(x_2)))).$
\\
The tree of the term $t$ is given on the Figure \ref{f1}:

\begin{figure}
  \includegraphics[width=10cm]{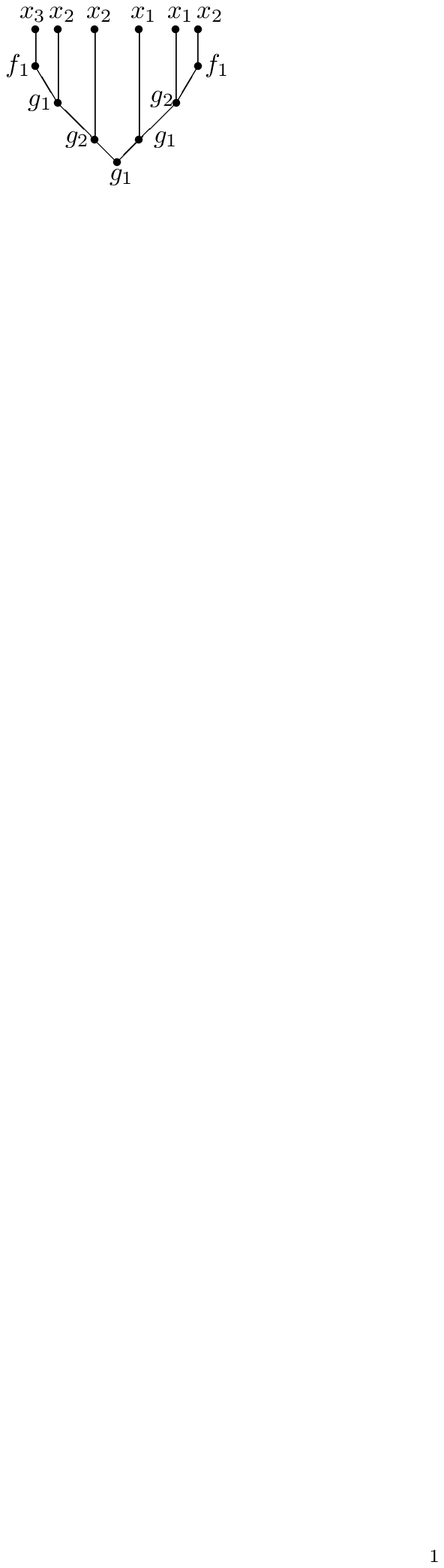}\\
  \caption{~~}\label{f1}
\end{figure}

The set of positions for $t$ is:
\\
$Pos(t)=\{\varepsilon,1,11,111,1111,12,112,2,21,22,221,222,2221\}$
and the corresponding subtrees to these positions are:
$t|_{1}=g_2(g_1(f_1(x_3),x_2),x_2),$\
$t|_{11}=g_1(f_1(x_3),x_2),$\
$t|_{12}=x_2,$\
$t|_{111}=f_1(x_3),$\
$t|_{112}=x_2,$\
$t|_{1111}=x_3,$ \
$t|_{2}=g_1(x_1,g_2(x_1,f_1(x_2))),$\
$t|_{21}=x_1,$\
$t|_{22}=g_2(x_1,f_1(x_2)),$\
$t|_{221}=x_1,$\
$t|_{222}=f_1(x_2),$\
$t|_{2221}=x_2.$
\\
There are eight possible assignments
and exactly six strong chains of subterms
which connect
the leaves of $t$ and  the root
of $t.$
\\
 It is easy to see that
$x_3\in Ess(t|_{k},{ A})$ for
$k=1111,111,11$ but
$ x_3\notin Ess(t|_{1},{ A}),$
and
$x_2\in Ess(t|_m,{ A})$ for
$ m=2221,222,22$ but
$ x_2\notin Ess(t|_{2},{ A}).$
If we apply the RFI-rule \ref{r}$(i)$ then the term $t$ can be reduced
to the term
\\
$t'=g_1(g_2(g_1(f_1(0),x_2),x_2),g_1(x_1,g_2(x_1,f_1(0)))).$
\\
This term is simpler than $t$ because of
$t\in W_{\tau}(X_3),$ but $t'\in W_{\tau}(X_2)$ and
$t\simeq_A t'.$
\\
To apply RFI-rule \ref{r}$(ii)$ let us note that
 for each $\gamma,\  \gamma\in Ass(X,{ F}_0)$
$${A}(\gamma,t|_{1})={ A}(\gamma,t|_{12}),\quad\mbox{\rm and}\quad
{A}(\gamma,t|_{2})={A}(\gamma,t|_{21}).$$
Thus we can obtain a reduction $t''$ of $t$ by replacing the subterms
$t|_{1}$ and $t|_{2}$ by subterms $t|_{12}$ and $t|_{21}.$
Hence,
$t''=g_1(t|_{12},t|_{21})=g_1(x_2,x_1).$
Clearly $t''$ is "much more" simple than $t$ and
$\forall \gamma\in Ass(X,{ F}_0)\quad
{ A}(\gamma,t)={ A}(\gamma,t'').$
\\
It is obvious, that the run of ${ A}$ over $\gamma$ and $t''$
will be more easy and more quick than over $\gamma$ and $t,$
but $t\simeq_{\A} t''.$
\end{e00}
\subsection{Optimal Automata-Languages}
Our next aim is to construct the "simplest" FTA ${ A}$
considered together with the tree-language
$L({ A})$
corresponding to ${ A}.$
 We have to pay attention on the
word "simplest" to avoid the conflict with the traditional understanding
of this notation.
\begin{d00}\label{d10}\rm
Let $t\in L({ A})$ be a tree recognizable by the automaton ${ A}.$
The tree $t$ is called {\it minimal} w.r.t. $A$ if\quad
$\forall s\in L({ A})\quad s\simeq_{ A} t\Ra (s\models_R t\quad
\mbox{\rm or}\quad s=t).$
\end{d00}
\begin{d00}\label{d11}\rm
A tree-language is called {\it minimal} w.r.t. $A$ if it consists only of minimal trees.
\end{d00}
\begin{d00}\label{d12}\rm
Two tree-languages $L_1$ and $L_2$ are called ${ A}-${\it equivalent}
$(L_1\sim L_2)$ iff
$$\forall t\in L_1\quad \exists s\in L_2\quad (t\simeq_{ A} s)\quad
\mbox{\rm and}\quad
\forall s\in L_2\quad \exists t\in L_1\quad (t\simeq_{ A} s).$$
\end{d00}
Let us consider the pair $\langle { A},L({ A})\rangle$ called
{\it automata-language}.
\\
It is important to compose such a pair with minimal components i.e. to find
such minimal automaton \cite{Com,Ges} which runs over minimal trees. Clearly
in this case the description of the automata-language is simplest, and such
pair will be called {\it optimal}.
\\
 There is a case when this task can be  solved.
\\
In \cite{Com,Ges} it is proved that the problem of finiteness of a
tree-language is decidable i.e. there exists an algorithm
$FA$ which
for each FTA
${ A}$ gives answer of the question:
Is the tree-language $L({ A})$ finite or no?
\\
Now we can describe a procedure for finding the optimal automata-language
when  a FTA ${ A}$ is given accepting finite tree-language $L(\A).$
\\
{\bf Procedure}
\begin{description}
\item[1] Use $FA$ to answer whether $L({ A})$ is finite or not?
\item[2] If $L({ A})$ is finite then use RFI-rules
to obtain minimal tree-language
$L_{min}\sim L({ A}).$
\item[3] Use an algorithm \cite{Com} to obtain a minimal FTA
${ A}_{min}$ which is equivalent to ${ A}$ i.e.
$L({ A}_{min})=L({ A})=L_{min}.$
\item[4] The pair $\langle { A}_{min},L_{min}\rangle$
is optimal.
\end{description}
An open problem is:
How to find the optimal automata-language (if it exists) when
 $L({ A})$ is not finite?
\par
There is an opportunity to
 describe some weaker conditions for essential
input variables which are fully sufficient for studying the  recognizable
tree languages.
\begin{d00}\rm\label{def22}
Let $t\in W_{\tau}(X)$ and let $A$
be a DFTA. An input variable $x_i\in Var(t)$ is called
{\it recognizably  essential (r-essential)} for the pair $(t,{ A})$
if there exist  two
assignments $\gamma_1, \gamma_2\in Ass(X,{ F}_0)$ such that
$$\gamma_1(x_i)\neq \gamma_2(x_i),\quad
\forall x_j\in X,\ j\neq i \quad \gamma_1(x_j)=\gamma_2(x_j)$$
 with
${ A}(\gamma_1,t)\in Q_f\iff { A}(\gamma_2,t)\notin Q_f.$
i.e. ${ A}$ stops in a
final state only with one of  $\gamma_1$ or
$\gamma_2.$
\end{d00}
The set of all $r-$essential inputs for $(t,{ A})$ is denoted by
$rEss(t,{ A}).$ The inputs from $Var(t)\setminus rEss(t,{ A})$
are called {\it r-fictive } for  $(t,{ A}).$
\\
Clearly, if $Q_f=Q$ or $Q_f=\emptyset$ then
$\forall t\in W_{\tau}(X)\quad rEss(t,{ A})=\emptyset.$
We will avoid such automaton as trivial case.
\\
The results for essential inputs may be proved
in the same way for $r-$essential ones. The notions for essential
variables may be introduced, too. For instance, the definition of
$\simeq_{rA}$ is:
\begin{d00}\rm\label{d91}
Two terms $t$ and $s$ are called {\it ${rA}$-equivalent}
$(t\simeq_{rA} s)$
iff
$$\forall  \gamma\in Ass(X,{ F}_0)\quad
{ A}(\gamma,t)\in Q_f\iff { A}(\gamma,s)\in Q_f.$$
\end{d00}
It is easy to see that:
\\ $(i)$\quad
If $\forall \gamma\in Ass(X,{ F}_0)\quad
({ A}(\gamma,t')\in Q_f\iff { A}(\gamma,t)\in Q_f)$
\\
then $rEss(t,{ A})=rEss(t',{ A}).$
\\$(ii)$\quad
If $t\in W_{\tau}(X)$ then
$rEss(t,{ A})\subset Ess(t,{ A}).$
\\ $(iii)$\quad
If the input $x_i$ is fictive for $t$ and ${ A}$ then $x_i$
is $r-$fictive for $t$ and ${ A}.$
\par
It is important that $A-$reductions in the case of $r-$fictive inputs are
stronger than in the case of usual fictive inputs,
 considered above  in Example  \ref{e1}.
\begin{e00}\label{e2}\rm
Let
${ A}=\langle Q,{ F}, Q_f,\Delta\rangle$  with
\\
${ F}_0=\{0,1,2\}$, ${ F}_1=\{f_0,f_1,f_2\}$, ${ F}_3=\{g_1,g_2\},$
$Q=\{q_0,q_1,q_2\}$, $Q_f=\{q_1,q_2\}$,
\\
$\Delta_0(0)=q_0$, $\Delta_0(1)=q_1$, $\Delta_0(2)=q_2$,
\\
$\Delta_1(f_i,q_j)=\left\{\begin{array}{ll}
q_1, & {\mbox{\rm if}}\quad i=j\\
q_0, & \mbox{\rm if} \quad i\neq j;
\end{array}
\right.$
\\
$\Delta_3(g_1,q_i,q_j,q_k)=q_m,\quad\mbox{\rm where}\quad m=i+j+k(mod\ 3),$
\\
$\Delta_3(g_2,q_i,q_j,q_k)=q_l,\quad\mbox{\rm where}\quad l=i.j.k(mod\ 3).$
\\
Let us consider the term
$t=g_2(f_2(x_1),f_2(x_2),g_1(f_0(x_3),f_1(x_3),g_2(1,1,x_3))).$ The
tree of $t$ is given at Figure \ref{f2}

\begin{figure}
  \includegraphics[width=12cm]{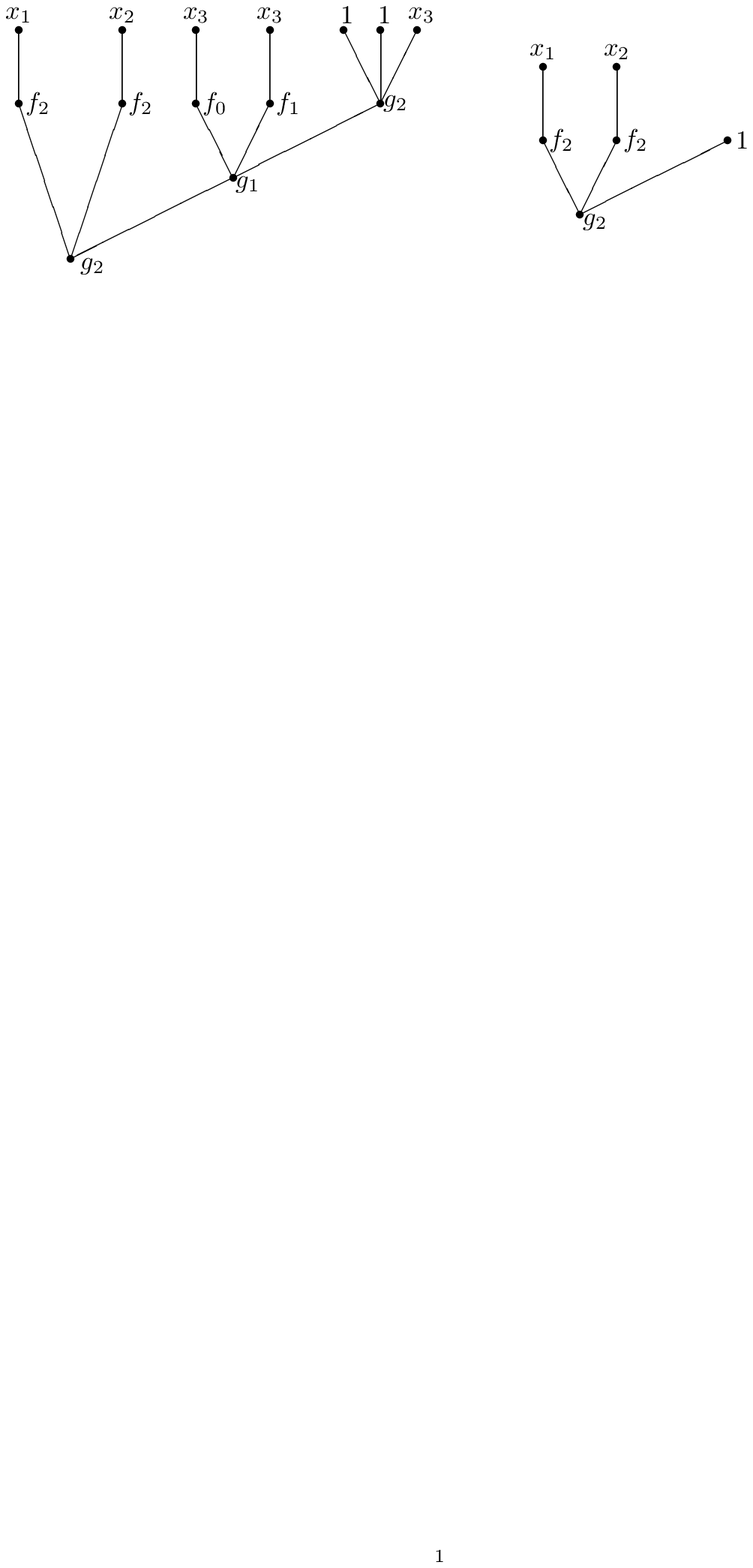}\\
  \caption{~~}\label{f2}
\end{figure}

It is easy to see that
 $x_3\in Ess(t,{ A})\setminus rEss(t,A)$ and $Ess(t,A)=\{x_1,x_2,x_3\}.$
Thus  we can't apply any RFI rules as above in Example \ref{e1}, but
if we use the fact that $x_3$ is $r-$fictive then the $rA-$reduction
of $t$ is possible and such reduction is given at Figure \ref{f2}.
Note that the trees at Figure \ref{f2}  are $rA-$equivalent.
\end{e00}

\end{document}